\documentclass[aps,pre,reprint,groupedaddress]{revtex4-1}
\usepackage{amssymb,amsfonts,amsmath}
\usepackage[dvips]{graphicx,color,psfrag}

\begin{document}
\title{Six Susceptible-Infected-Susceptible Models on Scale-free Networks}
\author{Satoru Morita}
\email[]{morita@sys.eng.shizuoka.ac.jp}
\affiliation{Department of Mathematical and Systems Engineering, Shizuoka University, Hamamatsu, 432-8561, Japan}
\date{\today}

\begin{abstract}
Spreading phenomena are ubiquitous in nature and society.
For example, disease, rumor, and information spread over 
underlying social and information networks.
It is well known that there is 
no threshold for epidemic models on scale-free networks;
this suggests that disease can spread on such networks,
regardless of  how low the contact rate may be.
In this paper, I consider 
six models with different contact and propagation mechanisms.
Each model is analyzed by degree-based mean-field theory.
I show that the presence or absence of an outbreak threshold
depends on the contact and propagation mechanism.
\end{abstract}
\pacs{87.10.Mn, 87.23.Cc, 89.65.Gh}

\maketitle

\section{Introduction}
The development of communication technology and transportation
has increased connectivity among people.
Outbreaks of several 
new infectious diseases, such as AIDS, SARS, swine flu, and Ebola,
have threatened human society.
These diseases spread over networks of contacts between individuals.
Computer viruses and worms spreading through the Internet
have caused severe economic damages all over the world.
Moreover, 
rumors, opinions and innovations spread through human networks.
Understanding the intrinsic mechanism 
behind spreading phenomena in networks is 
an important and urgent task\cite{pastor14,fu,barrat,newman10}.

Epidemic dynamics has often been described by ordinary differential equations,
that assume that the probability for an infected individual to encounter a susceptible host is uniform\cite{anderson,hethcote}.
However, social networks are not uniformly mixed
but are highly heterogeneous.
Many social networks, such as  telephone calls\cite{aiello},
e-mails\cite{ebel}, sexual relationships\cite{liljeros},
file actor coraboration\cite{amaral}, citation networks\cite{render}, and the Internet\cite{faloutsos},
have scale-free properties.
A network is called scale-free if the distribution of degree
(i.e., the number of links that connect to a node\cite{albert,barabasi,dorogovtsev,newman03})
obeys a power law:
\[
 P(k)\sim k^{-\gamma},
\]
where $k$ represents degree.
For most real world networks, the exponent $\gamma$ is between 2 and 3. 
High degree nodes are called hubs. 
Spreading processes in such networks have been intensively studied
recently\cite{pastor14,fu,barrat,newman10,pastor01a,pastor01b,parshani}.
It is well known that, 
for epidemic processes in scale-free networks,
the high heterogeneity of connections 
leads to the absence of an outbreak threshold\cite{pastor01a,pastor01b}.
In this paper, six different types of contact
and propagation dynamics on a network are considered.
Using degree-based mean-field theory, I show 
that the presence or absence of a threshold depends on the contacting 
and propagating dynamics.

In this paper, I adopt the susceptible-infected-susceptible (SIS) model.
The SIS model is one of the simplest models in epidemiology and is also known as the contact process model.
In an SIS model, a population with $N$ individuals
is categorize into two compartments: susceptible (S) and infected (I).
The disease is transmitted only when a susceptible individual 
is in contact with an infected individual. 
I assume that  $\lambda$ is the rate
of the contact  that is enough for the disease to be transmitted.
By rescaling time,
the recovery rate can be set to 1 without the loss of generality.
In the case of a fully mixed population, 
the model is represented by two stochastic events:
\begin{enumerate}
\item 
At the rate of $\lambda$, the contagious event is performed. 
An individual is chosen at random.
If the individual is susceptible, another individual is randomly chosen.
If the second individual is infected, the first individual 
becomes infected.
\item  
At the rate of 1, the recovery event is performed.
Choose an individual at random.
If the individual is infected, it recovers and becomes susceptible.
\end{enumerate}
Mean field theory shows that the average density of
infected individuals $\rho(t)$ follows the rate equation  
\begin{equation}
 \frac{d\rho(t)}{dt}=-\rho(t)+\lambda 
\rho(t)[1-\rho(t)].
\label{eq1}
\end{equation}
The equilibrium solution is calculated as
\begin{equation}
\rho^*= \left\{
\begin{array}{lcl}
\displaystyle 1-\frac{1}{\lambda} & \ & (\lambda \geq 1)\\
0 & \ &(\lambda\leq 1).
\end{array}
\right.
\label{eq00}
\end{equation}
There is an outbreak threshold $\lambda_c=1$,
above which epidemic spreads can occur.
Note that if the roles of susceptible and infected individuals 
are exchanged in the contagious event, result (\ref{eq00}) does not change.
In other words, the direction of the transmission is irrelevant
to the spreading phenomena in a fully mixed population. 
In the next section, I extend the model to include the network structure
and the direction of the transmission.

\section{Models and Analyses}
Here, six SIS models on networks are presented.
I assume that an individual is located at each node of a fixed network. 
The links of the network represent potential connections.
It is assumed that an individual is activated at each timestep.
Then, the active individual can contact its nearest neighbors on the network.
Two types of contacting mechanism are considered: 
(1) all neighbors or (2) only one neighbor is contacted at the same time.
In addition, two possibilities for transmission are considered:
an active individual is (a) the sender or (b) the receiver. 
I also consider the hybrid case (c): an active individual plays both roles.
By combining the contacting and  transmitting types,
six models are constructed, as shown as follows (see also table \ref{tab1}). 

\begin{table*}[tb]
\caption{Properties of six models.}
\begin{center}
  \begin{tabular}{ccccc}
    \hline
    model & contacts & active individual & outbreak threshold & equilibrium density of infected \\ 
    \hline \hline
    1a & all neighbors & sender & vanish & same as Pastor-Satorras and Vespignani's model\\
    \hline
    1b & all neighbors & receiver & vanish & lower than model 1a\\
    \hline
    1c & all neighbors & hybrid & vanish & intermediate of 1a and 1b\\
    \hline
    2a & one neighbor & sender & finite & lower than model 2b\\
    \hline
    2b & one neighbor & receiver & finite & same as well-mixed case\\
    \hline
    2c & one neighbor & hybrid & vanish & \\
    \hline
  \end{tabular}
\end{center}
\label{tab1}
\end{table*}

\subsection{Model 1a}
In model 1a, if an infected individual is activated,
all of its neighbors become infected. 
The contagious event in the previous section is replaced as follows
\begin{enumerate}
\item 
At the rate of $\lambda/\langle k \rangle$, the contagious event is performed. 
Choose an individual at random.
If it is infected, all of its neighbors get infected.
\end{enumerate}
Here, the contact rate $\lambda$ is divided by $\langle k \rangle$
because the average number of contact per unit time is set to 1.
Following the degree-based mean-field theory\cite{pastor01a,pastor01b},
consider the relative density $\rho_k(t)$ of infected individuals with degree $k$.
The rate equation for $\rho_k(t)$ is given by
\begin{equation}
  \frac{d\rho_k(t)}{dt}=-\rho_k(t)+ \frac{\lambda}{\langle k \rangle} 
[1-\rho_k(t)]k\Theta(t).
\label{eq2}
\end{equation}
The first term of the right-hand side is the recovery event.
The second term, which represents the contagious event, is 
proportional to the contact rate $\lambda/\langle k \rangle$,
the density of susceptible individuals $1-\rho_k(t)$,
the degree $k$ and the variable $\Theta(t)$, 
which is the probability that 
a link transmits the disease.
Since the probability that a link points to a node of degree $k$
equals $kP(k)/\langle k \rangle$, I obtain
\begin{equation}
 \Theta(t)=\frac{1}{\langle k \rangle}\sum_{k}kP(k)\rho_{k}(t).
\label{eq3}
\end{equation}
From Eq.~(\ref{eq2}), the equilibrium condition leads to
\begin{equation}
 \rho^*_k=\frac{k \lambda\Theta^*}{\langle k \rangle+k\lambda\Theta^*}.
\label{eq3b}
\end{equation}
The total density of the infected in the equilibrium state is determined as
\begin{equation}
\rho^*=\sum_k P(k) \rho_k^* 
\label{eq3d}
\end{equation}
Substituting Eq.~(\ref{eq3b}) into Eq.~(\ref{eq3}) and 
dividing both sides by $\Theta_*$, I obtain
\begin{equation}
1=
\displaystyle \frac{1}{\langle k \rangle}\sum_{k}kP(k)\
\frac{k\lambda}{\langle k \rangle+k\lambda\Theta^*}.
\label{eq3c}
\end{equation}
To calculate the outbreak threshold, $\lambda_c$,
I take the limit $\Theta^* \to 0$, and thus, I easily obtain
\begin{equation}
\lambda_c=
\frac{\langle k \rangle^2}{\langle k^2 \rangle}.
\label{eqlc1}
\end{equation}
Since the second moment of the degree $\langle k^2 \rangle$
diverges for scale-free networks with $\gamma\leq 3$, the outbreak
threshold can vanish, i.e., $\lambda_c=0$.
Thus, disease can spread on scale-free networks
no matter how low the contact rates may be.
This model is essentially the same as the
Pastor-Satorras and Vespignani model\cite{pastor01a,pastor01b}.

\subsection{Model 1b}
In model 1b, I reverse the direction of the contagious process of Model 1a.
An active susceptible individual gets infected
if there is at least one infected neighbor.
The contagious event is replaced as follows
\begin{enumerate}
\item 
At the rate of $\lambda/\langle k \rangle$, the contagious event is performed. 
Choose an individual at random.
If it is susceptible and has at least one infected neighbor, it becomes infected.
\end{enumerate}
In this case, the transmission probability in one time step is
\[
1-(1-\Theta(t))^k,
\]
rather than $k\Theta(t)$.
Therefore, the rate equation for $\rho_k(t)$ is rewritten as
\begin{equation}
  \frac{d\rho_k(t)}{dt}=-\rho_k(t)+ \frac{\lambda}{\langle k \rangle} 
[1-\rho_k(t)]\left[1-(1-\Theta(t))^k\right]
\label{eq2'}
\end{equation}
From Eq.~(\ref{eq2'}), the equilibrium condition leads to
\begin{equation}
 \rho^*_k=\frac{\lambda \left[1-(1-\Theta_*)^k\right]}
{\langle k \rangle+ \lambda\left[1-(1-\Theta_*)^k\right]}.
\label{eq3b'}
\end{equation}
Substituting Eq.~(\ref{eq3b'}) into Eq.~(\ref{eq3}), I obtain
a self consistent equation 
\begin{equation}
\Theta^*=
\displaystyle \frac{1}{\langle k \rangle}\sum_{k}kP(k)\
\frac{\lambda \left[1-(1-\Theta_*)^k\right]}
{\langle k \rangle+ \lambda\left[1-(1-\Theta_*)^k\right]}.
\label{eq3c'}
\end{equation}
Considering the limit $\Theta^* \to 0$, it is obvious that 
the outbreak threshold is given by Eq.~(\ref{eqlc1}) once more.
The absence of an outbreak threshold is seen as in model 1a.
Since $1-(1-\Theta(t))^k<k\Theta(t)$,
the equilibrium density $\rho_*$ is smaller than that for model 1a.

\subsection{Model 1c}
Next, I consider a hybrid of models 1a and 1b.
Thus, the contagious event is described as follows:
\begin{enumerate}
\item 
At the rate of $\lambda/2\langle k \rangle$, the contagious event is performed. 
Choose an individual at random.
If it is infected, all of its neighbors becomes infected.
If it is susceptible and has at least one infected neighbor, it becomes infected.
\end{enumerate}
In this case, the rate equation is formed as a compound of
Eqs.~(\ref{eq2}) and (\ref{eq2'}):
\begin{equation}
  \frac{d\rho_k(t)}{dt}=-\rho_k(t)+ \frac{\lambda}{2\langle k \rangle} 
[1-\rho_k(t)]\left[k\Theta(t)+ 1-(1-\Theta(t))^k\right]
\label{eq2''}
\end{equation}
In the same way as for model 1b,
I obtain a self consistent equation 
\begin{equation}
\Theta^*=
\displaystyle \frac{1}{\langle k \rangle}\sum_{k}kP(k)\
\frac{\lambda \left[k\Theta_*+1-(1-\Theta_*)^k\right]}
{2\langle k \rangle+ \lambda\left[k\Theta_*+1-(1-\Theta_*)^k\right]}.
\label{eq3c''}
\end{equation}
Thus, the outbreak threshold is given as Eq.~(\ref{eqlc1}).
The equilibrium density, $\rho_*$, is 
intermediate between those for models 1a and 1b.

\subsection{Model 2a}
In the above three models, it was assumed that 
an individual contacts all of its neighbors simultaneously.
Hereafter, I introduce three models
to restrict the contacts to only one.
In model 2a,
if an infected individual is activated,
only one neighbor can become infected.
The contagious event is replaced as follows.
\begin{enumerate}
\item 
At the rate of $\lambda$, the contagious event is performed. 
Choose an individual at random.
If the individual is infected, choose another individual among its neighbors randomly.
If the second individual is susceptible, it becomes infected.  
\end{enumerate}
In this case, the rate equation is rewritten as
\begin{equation}
 \frac{d\rho_k(t)}{dt}=-\rho_k(t)+\lambda 
[1-\rho_k(t)]\frac{k}{\langle k \rangle}\rho(t).
\label{eq5}
\end{equation}
Here, instead of $\Theta(t)$ in Eq.~({\ref{eq2}}) for model 1a, 
I need to use $\rho(t)$, which is  the probability that  an individual is infected:
\[
\rho(t)=\sum_{k}P(k)\rho_{k}(t).
\]
This comes from the fact that
the number of transmissions provided by an infected individual is 
proportional to the degree in model 1a, while
the degree is irrelevant to infectivity in model 2a.
The equilibrium condition leads to 
\begin{equation}
  \rho^*_k=\frac{k \lambda\rho^*}{\langle k \rangle+k\lambda\rho^*}.
\label{eq5b}
\end{equation}
Substituting Eq.~(\ref{eq5b}) into Eq.~(\ref{eq3d})
and dividing both sides by $\rho^*$, I obtain
\begin{equation}
 1=\sum_{k}P(k)\frac{k \lambda}{\langle k \rangle+k\lambda\rho^*}.
\label{eq5c}
\end{equation}
Taking  the limit as $\rho^*\to 0$, I find that
the outbreak threshold is
\[
 \lambda_c=1,
\]
regardless of the degree distribution $P(k)$.

\subsection{Model 2b}
Here, I reverse the direction of the contagious process of model 2a.
An active susceptible individual contacts only one neighbor;
if the neighbor is infected, the susceptible individual becomes infected.
The contagious event is replaced as follows.
\begin{enumerate}
\item 
At the rate of $\lambda$, the contagious event is performed. 
Choose an individual at random.
If the individual is susceptible, choose another individual among its neighbors randomly.
If the second individual is infected, the first individual becomes infected.
\end{enumerate}
In this case, the rate equation for $\rho_k(t)$ is written as
\begin{equation}
 \frac{d\rho_k(t)}{dt}=-\rho_k(t)+\lambda 
[1-\rho_k(t)]\Theta(t).
\label{eq4}
\end{equation}
Here, $k/\langle k\rangle$ is 
removed in Eq.~({\ref{eq2}}) for model 1a.
The equilibrium condition leads to
\begin{equation}
 \rho^*_k=\frac{ \lambda\Theta^*}{1+\lambda\Theta^*}.
\label{eq4b}
\end{equation}
In this case, $\rho^*_k$ is independent of the degree, $k$.
Thus, $\rho^*_k=\rho^*=\Theta^*$.
From eq.~(\ref{eq4b}), 
the nonzero solution $\rho^*>0$ satisfies 
\begin{equation}
1=\frac{\lambda}{1+\lambda \rho^*}.
\label{eq4c}
\end{equation}
Thus, the analytical solution is obtained as
\[
\rho^*=1-\frac{1}{\lambda}.
\]
The density of infected individuals $\rho^*$  coincides with SIS in a fully mixed population, regardless of the degree distribution.
As a result, the outbreak threshold is given by
\[\lambda_c=1.
\]
Jensen's inequality leads to
\[
\sum_{k}P(k)\frac{k \lambda}{\langle k \rangle+k\lambda\rho^*}
\geq\frac{\lambda}{1+\lambda \rho^*}.
\]
By comparing Eqs.~(\ref{eq5c}) and (\ref{eq4c}), I can deduce that 
in model 2b, the equilibrium density $\rho_*$ is larger than that in model 2a.

\subsection{Model 2c}
Finally, I consider a hybrid of models 2a and 2b.
Thus, the contagious event is given as follows
\begin{enumerate}
\item 
At the rate of $\lambda/2$, the contagious event is performed. 
Choose an individual at random and 
then choose another individual among its neighbors randomly.
If the first individual is susceptible and the second one is infected, 
the first one becomes infected.
If the first individual is infected and the second one is susceptible, 
the second one becomes infected. 
\end{enumerate}
In this case, the rate equation is formed as a compound of
eqs.~(\ref{eq4}) and (\ref{eq5}):
\begin{equation}
 \frac{d\rho_k(t)}{dt}=-\rho_k(t)
+\frac{\lambda}{2} \left[1-\rho_k(t)\right]
\left[\Theta(t)
+\frac{k}{\langle k \rangle}\rho(t)\right].
\label{eq6}
\end{equation}
The equilibrium condition leads to
\begin{equation}
 \rho_k=\frac{\lambda(\Theta^*+k\rho^*/\langle k \rangle )}
{2+\lambda(\Theta^*+k\rho^*/\langle k \rangle)}.
\label{eq6b}
\end{equation}
Substituting eq.~(\ref{eq6b}) into eqs.~(\ref{eq3}) and (\ref{eq3d}),
I obtain 
\[
\begin{array}{lcl}
\Theta^*&=&\displaystyle
 \frac{1}{\langle k \rangle}\sum_{k}kP(k)\
\frac{\lambda(\Theta^*+k\rho^*/\langle k \rangle)}
{2+\lambda(\Theta^*+k\rho^*/\langle k \rangle)},\\
\rho^*&=&\displaystyle\sum_{k}P(k)\
\frac{\lambda(\Theta^*+k\rho^*/\langle k \rangle)}
{2+\lambda(\Theta^*+k\rho^*/\langle k \rangle)}.
\end{array}
\]
To estimate the outbreak threshold, I take
the limit as $\Theta^* \to 0$ and $\rho^* \to 0$:
\begin{equation}
\begin{array}{lcl}
2 \Theta^* &=& \lambda_c(\Theta^*+\rho^*\langle k^2 \rangle/\langle k \rangle),\\
2 \rho^* &=& \lambda_c(\Theta^*+\rho^*).
\end{array}
\label{eq6c}
\end{equation}
Solving Eq.~(\ref{eq6c}), I obtain
\[
 \lambda_c=\frac{2}{1+\sqrt{\langle k^2 \rangle/\langle k \rangle}}
\]
For scale-free networks with exponent $\gamma<3$,
the outbreak threshold vanishes; i.e., $\lambda_c=0$.

\begin{figure*}[tb]
\begin{center}
\includegraphics[width=0.9\textwidth]{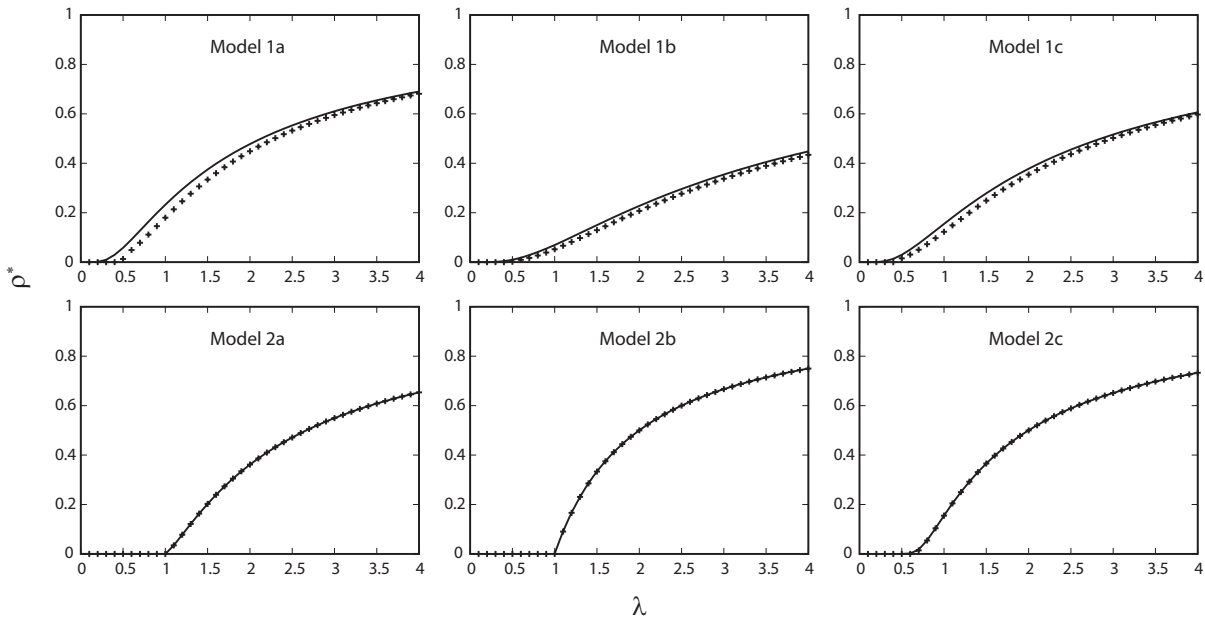}
\end{center}
\caption{The density of infected individuals $\rho^*$ is plotted as 
a function of $\lambda$ for the six different models. 
The curves show the theoretical predictions, while the crosses represent the numerical results.
The theoretical results are calculated from
Eqs.~(\ref{eq3c}), (\ref{eq3c'}), (\ref{eq3c''}), (\ref{eq5c}), (\ref{eq4c})
and (\ref{eq6c}) in the case of  Eq.~(\ref{eq0}). 
In the numerical simulations, the system size is set to $N=100000$ and
the average degree is set to $\langle k \rangle=4$.
Each point is obtained by averaging over 10000 unit time after 10000
relaxation time on 10 different network realizations. 
}
\label{fig}
\end{figure*}

\section{Numerical simulations}
To confirm the prediction made using  the degree-based mean-field 
theory, numerical simulations are performed.
Here, I adopt Barab\'{a}si and Albert's model (the BA model)\cite{barabasi}.
This model assumes that a new node is added to the network and 
that each new node is connected to $m\geq 1$
existing nodes with a probability that is proportional to the degree of the existing nodes.
An analytical approach based on the master equation 
shows that the degree distribution of the BA model is given by
\begin{equation}
P(k)=\frac{2m(m+1)}{k(k+1)(k+2)}  
\label{eq0}
\end{equation}
for $k\geq m$. Thus, for large values of $k$,
\[
 P(k)\sim k^{-3} .
\]
The average degree $\langle k \rangle$ is $2m$ asymptotically.
In fig.~1, the total density of the infected $\rho^*$ is 
plotted as a function of $\lambda$.
The crosses represent the
numerical results for  $N=100000$ and $m=2$, and
the curves show the theoretical solutions,
which are calculated using Eq.~(\ref{eq0}) as the degree distribution.
The theoretical calculations agree well with the numerical results.

\section{Discussion and Conclusion}
I have analyzed the spreading phenomena
on scale-free networks using 
six SIS models with different contact and propagation mechanisms.
The theoretical predictions showed good agreement with
the results of the numerical simulations.
The six models were divided into two types:
in the first type,
an active individual can contact all of its neighbors at the same time.
Here, the outbreak threshold can vanish,
regardless of the direction of propagation.
In the case where active individuals are senders (model 1a), 
the epidemic prevalence is larger than 
in the case where they are receivers (model 1b).
This is due to the fact that in model 1b,
propagation from more than one infected neighbors comes to nothing.
In the second type of model,
an active individual can contact only one neighbor.
In contrast to the first type, 
when the active individuals are receivers (model 2b),
the epidemic prevalence is larger than that for model 2a.
This result may look counterintuitive at first sight;
however, this result is not so surprising, 
because for model 2b,  an infected individual at a hub 
can transmit disease to more than one individual during a timestep,
in contrast to model 2a.
It may be more surprising that 
in the case where the transmission is bidirectional (model 2c),
the outbreak threshold can vanish,
while in the case the transmission is one way (models 2a and 2b),
the threshold remains finite.
These results are summarized in table \ref{tab1}.

In conclusion,
the epidemic prevalence depends entirely on the mechanisms of 
contact and propagation.
In fact, the six models handle
extreme situations and may not be realistic 
for practical contagious diseases.
However, the results of this paper have a wide-range of applications in 
the study of the spreading phenomena, not only for epidemic diseases
but also for such things as rumors and information.
This is expected to be a reference point 
when considering more complex models,
such as ones in which the network can change dynamically.

\begin{acknowledgments}
This work was supported by Grant-in-Aid for Scientific Research (No. 26400388) and CREST, JST.
Some of the numerical calculations were carried
out on machines at YITP of Kyoto University.
\end{acknowledgments}


\begin{thebibliography}{99}
\bibitem{pastor14}
R. Pastor-Satorras, C. Castellano, P. Van Mieghem, and A. Vespignani
``Epidemic processes in complex networks,''
arXiv:1408.2701, (2014).

\bibitem{fu}
X. Fu, M. Small, and G. Chen, 
\textit{Propagation Dynamics on Complex Networks: Models, Methods and Stability Analysis}, 
Higher Education Press, Beijing, (2014).

\bibitem{barrat}
A. Barrat, M. Barth\'{e}lemy, and A. Vespignani,
\textit{Dynamical Processes on Complex Networks}, 
Cambridge University Press, Cambridge, (2012).

\bibitem{newman10}
M. E. J. Newman,
\textit{Networks: An Introduction}, 
Oxford University Press, New York, (2010).

\bibitem{anderson}
R. M. Anderson and R. M. May
\textit{Infectious Diseases of
Humans: Dynamics and Control},
Oxford University Press, Oxford, (1991).

\bibitem{hethcote}
H. W. Hethcote, 
``The Mathematics of Infectious Diseases,''
SIAM Rev. {\bf 42}, 599, (2000).

\bibitem{aiello}
W. Aiello, F. Chung, and L. Lu,
``A random graph model for massive graphs in Proceedings''
the 32nd Annual ACM Symposium on Theory of Computing, 
pp.\ 171--180, 
(2000).

\bibitem{ebel}
K. Ebel, L. -I. Mielsch and S. Bornholdt, 
``Scale-free topology of e-mail networks,''
Phys. Rev. E {\bf 66}, 035103, (2002).

\bibitem{liljeros}
F. Liljeros, C. R. Edling, L. A. N. Amaral, H. E. Stanley and Y. Aberg, 
``The web of human sexual contacts,''
Nature {\bf 411}, 907, (2001).

\bibitem{amaral}
L. A. N. Amaral, A. Scala, M. Barth\'el\'emy, and 
H. E. Stanley, 
``Classes of small-world networks,''
Proc. Natl. Acad. Sci. USA {\bf 97}, 11149, (2000).


\bibitem{render}
S. Redner,
``How popular is your paper? An empirical study of the citation distribution,''
Eur. Phys. J. B {\bf 4}, 131, (1998).

\bibitem{faloutsos}
M. Faloutsos, P. Faloutsos and C. Faloutsos, 
``On power-law relationships of the internet topology,'' 
Computer Communications Review {\bf 29}, 251, (1999).

\bibitem{albert}
R. Albert and A. -L. Barab\'{a}si,
``Statistical mechanics of complex networks,''
Rev. Mod. Phys. {\bf 74}, 47, (2002).

\bibitem{barabasi}
A. -L. Barab\'{a}si and R. Albert,
``Emergence of scaling in random networks,''
Science {\bf 286}, 509, (1999).

\bibitem{dorogovtsev}
S. N. Dorogovtsev and J. F. F. Mendes,
\textit{Evolution of Networks},
Oxford University Press, New York, (2003).

\bibitem{newman03}
M. E. J. Newman,
``The structure and function of complex networks,''
SIAM Review {\bf 45}, 167, (2003).

\bibitem{pastor01a}
R. Pastor-Satorras and A. Vespignani,
``Epidemic Spreading in Scale-Free Networks,''
Phys. Rev. Lett. {\bf 86}, 3200, (2001).

\bibitem{pastor01b}
R. Pastor-Satorras and A. Vespignani, 
``Epidemic dynamics and endemic states in complex networks,''
Phys. Rev. E {\bf 63}, 066117, (2001).

\bibitem{parshani}
R. Parshani, S. Carmi, and S. Havlin
``Epidemic Threshold for the Susceptible-Infectious-Susceptible Model on Random Networks''
Phys. Rev. Lett. {\bf 104}, 25870, (2010).
\end{thebibliography}
\end{document}